\documentclass[12pt,prd,showpacs,tightenlines,nofootinbib]{revtex4}
\usepackage{bm}
\usepackage{graphics}
\usepackage{rotating}
\usepackage{epsfig}
\begin{document}
\title{\begin{flushright}{\rm\normalsize HU-EP-05/82}\end{flushright}
Masses of heavy tetraquarks in the relativistic quark model}
\author{D. Ebert}
\affiliation{Institut f\"ur Physik, Humboldt--Universit\"at zu Berlin,
Newtonstr. 15, D-12489  Berlin, Germany}
\author{R. N. Faustov}
\author{V. O. Galkin}
\affiliation{Institut f\"ur Physik, Humboldt--Universit\"at zu Berlin,
Newtonstr. 15, D-12489 Berlin, Germany}
\affiliation{Dorodnicyn Computing Centre, Russian Academy of Sciences,
  Vavilov Str. 40, 119991 Moscow, Russia}
\begin{abstract}
The masses of heavy tetraquarks with hidden charm and bottom are
calculated in the framework of the relativistic quark model. The
tetraquark is considered as the bound state of a heavy-light diquark
and   
antidiquark. The light quark in a heavy-light diquark is treated
completely relativistically. The internal structure of the diquark is
taken into account by calculating the diquark-gluon form
factor in terms of the diquark wave functions. New experimental
data on charmonium-like states above open charm threshold are discussed.
The obtained results indicate that the $X(3872)$ can be the 
tetraquark state  with hidden charm. The masses of ground state
tetraquarks  with hidden bottom  are found to be below the open bottom
threshold.     
\end{abstract}


\maketitle

Recently,
significant progress in experimental investigations of
charmonium spectroscopy has been achieved. Several new states
$X(3872)$, $Z(3931)$, $Y(3943)$, $X(3943)$, 
$Y(4260)$ were observed
\cite{exp} 
which provide challenge to the theory, since not all 
of them can be easily accommodated as the  $c\bar c$-charmonium. 
The most natural conjecture is that these states are the
multiquark composite systems considered long ago e.g. in
\cite{jm}. The proposal to revisit the multiquark picture
using diquarks has been raised by Jaffe and Wilczek \cite{jw}.
Currently the best established state is the 
narrow $X(3872)$ which was originally discovered in $B$ decays
\cite{belle,babar} and later confirmed in $p\bar p$ collisions
\cite{cdf,d0}. Its mass and observed decays, which favour $J^{PC}=1^{++}$
assignment, make a
$c\bar c$ interpretation problematic \cite{elq}.   
Different theoretical interpretations of the $X(3872)$ state were put
forward which use the near proximity of its mass to the $D^0\bar
D^{*0}$ threshold. The most popular ones are: the $D^0-\bar D^{*0}$
molecular state bound by pion and quark exchanges \cite{molec}; an
$s$-wave cusp at $D^0\bar D^{*0}$ threshold \cite{bugg} and the 
diquark-antidiquark $[cq][\bar c\bar q]$ tetraquark state \cite{mppr}
($q=u,d$). 
  
Maiani et al. \cite{mppr}  in the framework of the phenomenological
constituent quark 
model considered the masses of hidden charm diquark-antidiquark states
in terms of the constituent diquark mass and spin-spin 
interactions. They identified the $X(3872)$ with the $S$-wave bound state
of a
spin one and spin zero diquark and antidiquark with the symmetric
diquark-spin distribution $([cq]_{S=1}[\bar c\bar q]_{S=0}+[cq]_{S=0}[\bar c\bar
q]_{S=1})$ and used its mass to fix the constituent diquark
mass. Spin-spin couplings were fixed from the analysis of the observed
meson and baryon masses. On this basis they predicted the existence of
a $2^{++}$ state $[cq]_{S=1}[\bar c\bar q]_{S=1}$ that can be
associated to the $Y(3943)$. They also 
argued  \cite{mppr1} that $Y(4260)$ could be the first orbital
excitation of the charm-strange diquark-antidiquark state 
$([cs]_{S=0}[\bar c \bar s]_{S=0})_{P-{\rm wave}}$.  In
Ref.~\cite{bmppr} it is pointed out that non-leptonic $B$ decays
provide a favourable environment for the production of hidden charm
diquark-antidiquark bound states. In contrast it is argued \cite{suz}
that the observed  $X(3872)$ production in $B$ decays and in
high-energy $p\bar p$ collisions is too large for a loosely bound molecule
(with binding energy of 1 MeV or less).  

In this paper we use the relativistic quark model \cite{efg,egf} based
on the quasipotential approach to calculate the mass
spectra of tetraquarks with hidden charm and bottom as the
heavy-light diquark-antidiquark bound states ($[Qq][\bar Q\bar q]$,
$Q=c,b$). Recently we considered the mass spectra of doubly heavy
($QQq$) \cite{efgm} and heavy ($qqQ$) \cite{hbar} baryons in the
heavy-diquark--light-quark and light-diquark--heavy-quark
approximations, respectively. The light quarks and light diquarks were
treated completely relativistically. The internal structure of the light
and heavy diquarks was taken into account by calculating diquark-gluon
form factors on the basis of the determined diquark wave
functions. The found good agreement \cite{hbar} with available
experimental data gives additional motivation for considering
diquarks as  
reasonable building blocks of hadrons. It is important to
note that all parameters of our model were determined from the
previous considerations of meson mass spectra and decays, and we will
keep them fixed in the following analysis of heavy tetraquarks.

In the quasipotential approach and diquark-antidiquark picture of
heavy tetraquarks the interaction of two quarks in a diquark and the heavy
diquark-antidiquark interaction in a tetraquark are described by the
diquark wave function ($\Psi_{d}$) of the bound quark-quark state
and by the tetraquark wave function ($\Psi_{T}$) of the bound
diquark-antidiquark state respectively,  which satisfy the
quasipotential equation of the Schr\"odinger type \cite{efg}
\begin{equation}
\label{quas}
{\left(\frac{b^2(M)}{2\mu_{R}}-\frac{{\bf
p}^2}{2\mu_{R}}\right)\Psi_{d,T}({\bf p})} =\int\frac{d^3 q}{(2\pi)^3}
 V({\bf p,q};M)\Psi_{d,T}({\bf q}),
\end{equation}
where the relativistic reduced mass is
\begin{equation}
\mu_{R}=\frac{E_1E_2}{E_1+E_2}=\frac{M^4-(m^2_1-m^2_2)^2}{4M^3},
\end{equation}
and $E_1$, $E_2$ are given by
\begin{equation}
\label{ee}
E_1=\frac{M^2-m_2^2+m_1^2}{2M}, \quad E_2=\frac{M^2-m_1^2+m_2^2}{2M},
\end{equation}
here $M=E_1+E_2$ is the bound state mass (diquark or tetraquark),
$m_{1,2}$ are the masses of quarks ($q_1$ and $q_2$) which form
the diquark or of the diquark ($d$) and antiquark ($d'$) which form
the heavy tetraquark ($T$), and ${\bf p}$  is their relative momentum.  
In the center of mass system the relative momentum squared on mass shell 
reads
\begin{equation}
{b^2(M) }
=\frac{[M^2-(m_1+m_2)^2][M^2-(m_1-m_2)^2]}{4M^2}.
\end{equation}

The kernel 
$V({\bf p,q};M)$ in Eq.~(\ref{quas}) is the quasipotential operator of
the quark-quark or diquark-antidiquark interaction. It is constructed with
the help of the
off-mass-shell scattering amplitude, projected onto the positive
energy states. In the following analysis we closely follow the
similar construction of the quark-antiquark interaction in mesons
which were extensively studied in our relativistic quark model
\cite{efg,egf}. For
the quark-quark interaction in a diquark we use the relation
$V_{qq}=V_{q\bar q}/2$ arising under the assumption about the octet
structure of the interaction  from the difference in the $qq$ and
$q\bar q$  colour states.\footnote{
It is important to study diquark correlations in
gauge-invariant color-singlet hadron states on the
lattice.} 
An important role in this construction is
played by the Lorentz-structure of the confining  interaction. 
In our analysis of mesons while  
constructing the quasipotential of the quark-antiquark interaction, 
we adopted that the effective
interaction is the sum of the usual one-gluon exchange term with the
mixture 
of long-range vector and scalar linear confining potentials, where
the vector confining potential contains the Pauli terms.  
We use the same conventions for the construction of the quark-quark
and diquark-antidiquark interactions in the tetraquark. The
quasipotential  is then defined  as follows \cite{efgm,egf} 

(a) for the quark-quark ($Qq$) interaction
 \begin{equation}
\label{qpot}
V({\bf p,q};M)=\bar{u}_{1}(p)\bar{u}_{2}(-p){\cal V}({\bf p}, {\bf
q};M)u_{1}(q)u_{2}(-q),
\end{equation}
with
\[
{\cal V}({\bf p,q};M)=\frac12\left[\frac43\alpha_sD_{ \mu\nu}({\bf
k})\gamma_1^{\mu}\gamma_2^{\nu}+ V^V_{\rm conf}({\bf k})
\Gamma_1^{\mu}({\bf k})\Gamma_{2;\mu}(-{\bf k})+
 V^S_{\rm conf}({\bf k})\right],
\]
Here  $\alpha_s$ is the QCD coupling constant,  $D_{\mu\nu}$ is the 
gluon propagator in the Coulomb gauge
\begin{equation}
D^{00}({\bf k})=-\frac{4\pi}{{\bf k}^2}, \quad D^{ij}({\bf k})=
-\frac{4\pi}{k^2}\left(\delta^{ij}-\frac{k^ik^j}{{\bf k}^2}\right),
\quad D^{0i}=D^{i0}=0,
\end{equation}
and ${\bf k=p-q}$; $\gamma_{\mu}$ and $u(p)$ are 
the Dirac matrices and spinors
\begin{equation}
\label{spinor}
u^\lambda({p})=\sqrt{\frac{\epsilon(p)+m}{2\epsilon(p)}}
\left(\begin{array}{c}
1\\ \displaystyle\frac{\mathstrut\bm{\sigma}{\bf p}}
{\mathstrut\epsilon(p)+m}
\end{array}\right)
\chi^\lambda,
\end{equation}
with $\epsilon(p)=\sqrt{{\bf p}^2+m^2}$. 

The effective long-range vector vertex of the quark is
defined \cite{egf} by
\begin{equation}
\Gamma_{\mu}({\bf k})=\gamma_{\mu}+
\frac{i\kappa}{2m}\sigma_{\mu\nu}\tilde k^{\nu}, \qquad \tilde
k=(0,{\bf k}),
\end{equation}
where $\kappa$ is the Pauli interaction constant characterizing the
anomalous chromomagnetic moment of quarks. In the configuration space
the vector and scalar confining potentials in the nonrelativistic
limit reduce to
\begin{eqnarray}
V^V_{\rm conf}(r)&=&(1-\varepsilon)V_{\rm conf}(r),\nonumber\\
V^S_{\rm conf}(r)& =&\varepsilon V_{\rm conf}(r),
\end{eqnarray}
with 
\begin{equation}
V_{\rm conf}(r)=V^S_{\rm conf}(r)+
V^V_{\rm conf}(r)=Ar+B,
\end{equation}
where $\varepsilon$ is the mixing coefficient.

(b) for diquark-antidiquark ($d\bar d'$) interaction
\begin{eqnarray}
\label{dpot}
V({\bf p,q};M)&=&\frac{\langle d(P)|J_{\mu}|d(Q)\rangle}
{2\sqrt{E_dE_d}}  \frac43\alpha_SD^{ \mu\nu}({\bf 
k})\frac{\langle d'(P')|J_{\nu}|d'(Q')\rangle}
{2\sqrt{E_{d'}E_{d'}}} \cr
&&+\psi^*_d(P)\psi^*_{d'}(P')\left[J_{d;\mu}J_{d'}^{\mu}
V_{\rm conf}^V({\bf k})+V^S_{\rm conf}({\bf k})\right]\psi_d(Q)\psi_{d'}(Q'), 
\end{eqnarray}
where $\langle
d(P)|J_{\mu}|d(Q)\rangle$ is the vertex of the 
diquark-gluon interaction which takes into account the finite size of
the diquark and is discussed 
below 
$\Big[$$P^{(')}=(E_{d^{(')}},\pm{\bf p})$ and 
$Q^{(')}=(E_{d^{(')}},\pm{\bf q})$,
$E_d=(M^2-M_{d'}^2+M_d^2)/(2M)$ and $E_{d'}=(M^2-M_d^2+M_{d'}^2)/(2M)$ 
$\Big]$.

The diquark state in the confining part of the diquark-antidiquark
quasipotential (\ref{dpot}) is described by the wave functions
\begin{equation}
  \label{eq:ps}
  \psi_d(p)=\left\{\begin{array}{ll}1 &\qquad \text{ for scalar diquark}\\
\varepsilon_d(p) &\qquad \text{ for axial vector diquark}
\end{array}\right. ,
\end{equation}
where the four vector
\begin{equation}\label{pv}
\varepsilon_d(p)=\left(\frac{(\bm{\varepsilon}_d {\bf
p})}{M_d},\bm{\varepsilon}_d+ \frac{(\bm{\varepsilon}_d {\bf p}){\bf
  p}}{M_d(E_d(p)+M_d)}\right), \qquad \varepsilon^\mu_d(p) p_\mu=0,  
\end{equation} 
is the polarization vector of the axial vector
diquark with momentum ${\bf p}$, $E_d(p)=\sqrt{{\bf p}^2+M_d^2}$ and
$\varepsilon_d(0)=(0,\bm{\varepsilon}_d)$ is the polarization vector in
the diquark rest frame. The effective long-range vector vertex of the
diquark can be presented in the form  
\begin{equation}
  \label{eq:jc}
  J_{d;\mu}=\left\{\begin{array}{ll}
  \frac{\displaystyle (P+Q)_\mu}{\displaystyle
  2\sqrt{E_dE_d}}&\qquad \text{ for scalar diquark}\cr
-\; \frac{\displaystyle (P+Q)_\mu}{\displaystyle2\sqrt{E_dE_d}}
  +\frac{\displaystyle i\mu_d}{\displaystyle 2M_d}\Sigma_\mu^\nu 
\tilde k_\nu
  &\qquad \text{ for axial 
  vector diquark}\end{array}\right. ,
\end{equation}
where $\tilde k=(0,{\bf k})$. Here the antisymmetric tensor
\begin{equation}
  \label{eq:Sig}
  \left(\Sigma_{\rho\sigma}\right)_\mu^\nu=-i(g_{\mu\rho}\delta^\nu_\sigma
  -g_{\mu\sigma}\delta^\nu_\rho)
\end{equation}
and the axial vector diquark spin ${\bf S}_d$ is given by
$(S_{d;k})_{il}=-i\varepsilon_{kil}$; $\mu_d$ is the total
chromomagnetic moment of the axial vector 
diquark.

The constituent quark masses $m_b=4.88$ GeV, $m_c=1.55$ GeV,
$m_u=m_d=0.33$ GeV, $m_s=0.5$ GeV and 
the parameters of the linear potential $A=0.18$ GeV$^2$ and $B=-0.3$ GeV
have the usual values of quark models.  The value of the mixing
coefficient of vector and scalar confining potentials $\varepsilon=-1$
has been determined from the consideration of charmonium radiative
decays \cite{efg} and the heavy quark expansion \cite{fg}. 
The universal Pauli interaction constant $\kappa=-1$ has been
fixed from the analysis of the fine splitting of heavy quarkonia ${
}^3P_J$- states \cite{efg}. In this case the long-range chromomagnetic
interaction of quarks vanishes in accord with the flux tube
model. Finally, we choose the total
chromomagnetic moment of the axial vector diquark $\mu_d=0$.\footnote{ In 
  our previous papers \cite{efgm,hbar} on heavy baryons we 
  adopted the value $\mu_d=2$ following the 
analogy with QED, which might
  be valid only for the perturbative one-gluon exchange part of the quark
  potential.  However, in the above cases
 the long-range chromomagnetic
  diquark moment always contributed being multiplied by the 
vanishing total
  chromomagnetic long-range quark moment and, thus its
  value was irrelevant for the numerical results.} Such a
choice appears to be natural, since the long-range
chromomagnetic interaction of diquarks proportional to  $\mu_d$ 
then also vanishes  in accord with the flux tube model.

At a first step, we calculate the masses and form factors of the
heavy-light 
diquarks. As it is well known, the light quarks are highly
relativistic, which makes the $v/c$ expansion inapplicable and thus,
a completely relativistic treatment is required. To achieve this goal in
describing heavy-light 
diquarks, we closely follow our recent consideration of the spectra of
light diquarks in heavy baryons and adopt the same procedure to make
the relativistic quark potential local by replacing
$\epsilon_{1,2}(p)\equiv\sqrt{m_{1,2}^2+{\bf p}^2}\to E_{1,2}$  
(see discussion in Ref.~\cite{lmes}). The resulting
light-quark--heavy-quark interaction potential is the same as the
light quark-quark interaction \cite{hbar} and is equal to 1/2 of the
$Q\bar q$ interaction in the heavy-light meson.\footnote{The masses of
  the ground state heavy-light mesons are well reproduced with this
  $Q\bar q$ potential.}  We solve numerically the quasipotential
equation with this complete relativistic potential which depends on the
diquark mass in a complicated highly nonlinear way. The obtained
ground-state masses of scalar and axial vector heavy-light diquarks
are presented in Table~\ref{tab:dmass}.

\begin{table}
  \caption{Masses $M$ and form factor  parameters of heavy-light
    diquarks. $S$ and $A$ 
    denote scalar and axial vector diquarks antisymmetric $[Q,q]$ and
    symmetric $\{Q,q\}$ in flavour, respectively. }
  \label{tab:dmass}
\begin{ruledtabular}
\begin{tabular}{cccccccc}
Quark& Diquark&  
\multicolumn{3}{l}{\underline{\hspace{2.5cm}$Q=c$\hspace{2.5cm}}}
\hspace{-3.4cm}
&\multicolumn{3}{l}{\underline{\hspace{2.5cm}$Q=b$\hspace{2.5cm}}}
\hspace{-3.4cm} \\
content &type & $M$ (MeV)&$\xi$ (GeV)&$\zeta$ (GeV$^2$)  & $M$
(MeV)&$\xi$ (GeV)&$\zeta$ (GeV$^2$) \\
\hline
$[Q,q]$& $S$ & 1973& 2.55 &0.63 & 5359 &6.10 & 0.55 \\
$\{Q,q\}$& $A$ & 2036& 2.51 &0.45 & 5381& 6.05 &0.35 \\
$[Q,s]$ & $S$& 2091& 2.15 & 1.05 & 5462 & 5.70 &0.35 \\
$\{Q,s\}$& $A$ & 2158&2.12& 0.99 & 5482 & 5.65 &0.27
  \end{tabular}
\end{ruledtabular}
\end{table}

In order to determine the diquark interaction with the gluon field, which
takes into account the diquark structure, it is
necessary to calculate the corresponding matrix element of the quark
current between diquark states. This diagonal matrix element can be
parameterized by the following set of elastic form factors

(a) scalar diquark ($S$)
\begin{equation}
  \label{eq:sff}
  \langle S(P)\vert J_\mu \vert S(Q)\rangle=h_+(k^2)(P+Q)_\mu,
\end{equation}

(b) axial vector diquark ($A$) 
\begin{eqnarray}
  \label{eq:avff}
\langle A(P)\vert J_\mu \vert A(Q)\rangle&=&
-[\varepsilon_d^*(P)\cdot\varepsilon_d(Q)]h_1(k^2)(P+Q)_\mu\cr
&&+h_2(k^2)
\left\{[\varepsilon_d^*(P) \cdot Q]\varepsilon_{d;\mu}(Q)+
  [\varepsilon_d(Q) \cdot P] 
\varepsilon^*_{d;\mu}(P)\right\}\cr
&&+h_3(k^2)\frac1{M_{A}^2}[\varepsilon^*_d(P) \cdot Q]
    [\varepsilon_d(Q) \cdot P](P+Q)_\mu, 
\end{eqnarray}
where $k=P-Q$ and $\varepsilon_d(P)$ is the polarization vector of the
axial vector diquark (\ref{pv}).

\begin{figure}
\centerline{\includegraphics[height=12cm,angle=-90]{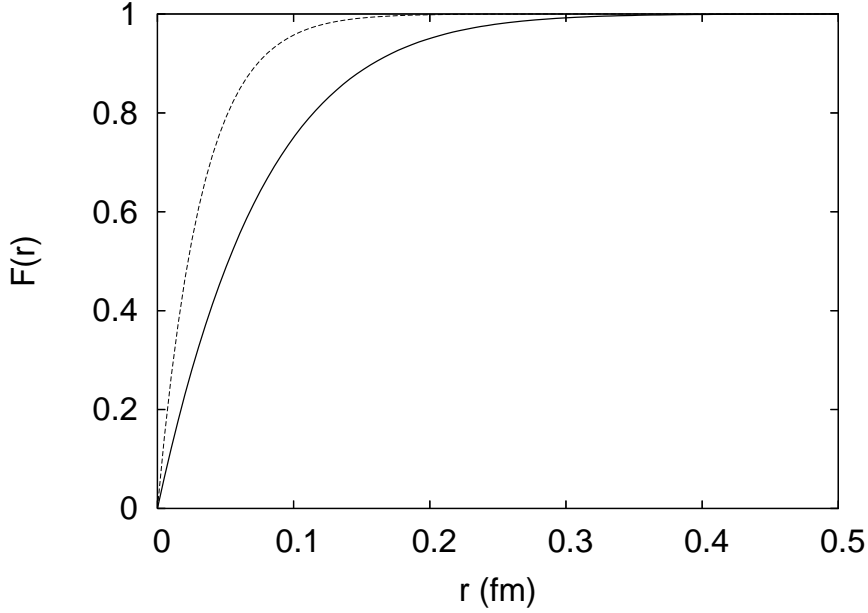}}
  \caption{\label{fig:ff} The form factors $F(r)$ for  $\{c,q\}$
    (solid line) and $\{b,q\}$ (dashed line) axial vector
    diquarks.}
\end{figure}

Using the quasipotential approach with the impulse approximation for the
vertex function of the quark-gluon interaction we find \cite{hbar} 
\begin{eqnarray*}
  h_+(k^2)&=&h_1(k^2)=h_2(k^2)=F({\bf k}^2),\cr
h_3(k^2)&=&0,
\end{eqnarray*}
\begin{eqnarray}\label{eq:hf}
F({\bf k}^2)&=&\frac{\sqrt{E_{d}M_{d}}}{E_{d}+M_{d}}
  \int \frac{d^3p}{(2\pi )^3} \bar\Psi_{d}
\left({\bf p}+
\frac{2\epsilon_{2}(p)}{E_{d}+M_{d}}{\bf k } \right)
\sqrt{\frac{\epsilon_1(p)+m_1}{\epsilon_1(p+k)+m_1}}
\Biggl[\frac{\epsilon_1(p+k)+\epsilon_1(p)}
{2\sqrt{\epsilon_1(p+k)\epsilon_1(p)}}\cr
&&+\frac{\bf p k}{2\sqrt{\epsilon_1(p+k)\epsilon_1(p)}
(\epsilon_1(p)+m_1)} \Biggr]\Psi_{d}({\bf
  p})+(1\leftrightarrow 2),
\end{eqnarray}
where $\Psi_{d}$ are the diquark wave functions.
We calculated the corresponding form factors $F(r)/r$ which are the Fourier
transforms of $F({\bf k}^2)/{\bf k}^2$ using the diquark wave
functions found 
by numerical solving the quasipotential equation.
Our estimates show that this form factor can be approximated  with a
high accuracy by the expression 
\begin{equation}
  \label{eq:fr}
  F(r)=1-e^{-\xi r -\zeta r^2},
\end{equation}
which agrees with previously used approximations \cite{efgm}.
The values of parameters $\xi$ and $\zeta$ for heavy-light
scalar diquark 
$[Q,q]$ and axial vector 
diquark
$\{Q,q\}$ ground states are
given in Table~\ref{tab:dmass}. In Fig.~\ref{fig:ff} we plot the functions
$F(r)$ for  $\{Q,q\}$ axial vector diquarks.

At a second step, we calculate the masses of heavy tetraquarks 
considered as the
bound state of a heavy-light diquark and antidiquark.
For  the potential of the heavy diquark-antidiquark  interaction
(\ref{dpot}) we get 
\begin{eqnarray}
  \label{eq:pot}
  V(r)&=&V_{\rm Coul}(r)+V_{\rm conf}(r)+\frac1{E_1E_2}\Biggl\{{\bf
    p}\left[V_{\rm Coul}(r)+V^V_{\rm conf}(r)\right]{\bf p} -\frac14 
\Delta V^V_{\rm conf}(r)+ V'_{\rm Coul}(r)\frac{{\bf
    L}^2}{2r}\cr
&&
+\frac1{r}\left[V'_{\rm Coul}(r)+\frac{\mu_d}4\left(\frac{E_1}{M_1}
+\frac{E_2}{M_2}\right)V'^V_{\rm conf}(r)\right]{\bf L}
({\bf S}_1+{\bf S}_2)\cr
&&
+\frac{\mu_d}4\left(\frac{E_1}{M_1}
-\frac{E_2}{M_2}\right)\frac{V'^V_{\rm conf}(r)}{r}{\bf L}
({\bf S}_1-{\bf S}_2)\cr
&&
+\frac13\left[\frac1{r}{V'_{\rm Coul}(r)}-V''_{\rm Coul}(r)
+\frac{\mu_d^2}4\frac{E_1E_2}{M_1M_2}
\left(\frac1{r}{V'^V_{\rm conf}(r)}-V''^V_{\rm
    conf}(r)\right)\right]\left[\frac3{r^2}({\bf S}_1{\bf r}) ({\bf
  S}_2{\bf r})- 
{\bf S}_1{\bf S}_2\right]\cr
&&
+\frac23\left[\Delta V_{\rm Coul}(r)+\frac{\mu_d^2}4\frac{E_1E_2}{M_1M_2}
\Delta V^V_{\rm conf}(r)\right]{\bf S}_1{\bf S}_2\Biggr\},
\end{eqnarray}
where 
$$V_{\rm Coul}(r)=-\frac{4}{3}\alpha_s \frac{F_1(r)F_2(r)}{r}$$ 
is the
Coulomb-like one-gluon exchange potential which takes into
account the finite sizes of the diquark and antidiquark through
corresponding form factors $F_{1,2}(r)$. Here ${\bf S}_{1,2}$ and ${\bf
  L}$ are spin operators of the
diaquark and antidiquark and their orbital momentum.

\begin{table}
  \caption{Masses of charm diquark-antidiquark states (in MeV). $S$ and $A$
    denote scalar and axial vector diquarks. }
  \label{tab:cmass}
\begin{ruledtabular}
\begin{tabular}{ccccc}
State& Diquark &
\multicolumn{3}{l}{\underline{\hspace{3.6cm}Mass\hspace{3.6cm}}} 
\hspace{-5.5cm} \\
$J^{PC}$ & content& $cq\bar c\bar q$ &$cs\bar c\bar s$ & $cs\bar c\bar
q/ cq\bar c\bar s$ \\
\hline
$1S$\\
$0^{++}$ & $S\bar S$ & 3812 & 4051 & 3922\\
$1^{+\pm}$ & $(S\bar A\pm \bar S A)/\sqrt2$& 3871& 4113 & 3982\\
$0^{++}$& $A\bar A$ & 3852 & 4110& 3967\\
$1^{+-}$& $A\bar A$ & 3890 & 4143& 4004\\
$2^{++}$& $A\bar A$ & 3968 & 4209&4080\\
$1P$\\
$1^{--}$& $S\bar S$ &4244 & 4466 & 4350
 \end{tabular}
\end{ruledtabular}
\end{table}

\begin{table}
  \caption{Masses of bottom diquark-antidiquark states (in MeV). $S$ and $A$
    denote scalar and axial vector diquarks. }
  \label{tab:bmass}
\begin{ruledtabular}
\begin{tabular}{ccccc}
State& Diquark &
\multicolumn{3}{l}{\underline{\hspace{3.7cm}Mass\hspace{3.7cm}}} 
\hspace{-5.5cm} \\
$J^{PC}$ & content& $bq\bar b\bar q$ &$bs\bar b\bar s$ & $bs\bar b\bar
q/ bq\bar b\bar s$ \\
\hline
$1S$\\
$0^{++}$ & $S\bar S$ & 10471 & 10662 & 10572\\
$1^{+\pm}$ & $(S\bar A\pm \bar S A)/\sqrt2$& 10492& 10682 & 10593\\
$0^{++}$& $A\bar A$ & 10473 & 10671& 10584\\
$1^{+-}$& $A\bar A$ & 10494 & 10686& 10599\\
$2^{++}$& $A\bar A$ & 10534 & 10716& 10628\\
$1P$\\
$1^{--}$& $S\bar S$ & 10807 & 11002 & 10907
 \end{tabular}
\end{ruledtabular}
\end{table}

\begin{table}
  \caption{Thresholds for open charm decays and nearby hidden-charm
    thresholds.} 
  \label{tab:cthr}
\begin{ruledtabular}
\begin{tabular}{cccccc}
Channel& Threshold (MeV)&Channel& Threshold (MeV)&Channel& Threshold
(MeV)\\ 
\hline
$D^0\bar D^0$& 3729.4 &$D_s^+ D_s^-$& 3936.2&$D^0 D_s^\pm$& 3832.9\\
$D^+D^-$& 3738.8& $\eta' J/\psi$& 4054.7& $D^\pm D_s^\mp$ & 3837.7\\
$D^0\bar D^{*0}$ & 3871.3& $D_s^\pm D_s^{*\mp}$& 4080.0&$D^{*0} D_s^\pm$ &
3975.0\\
$\rho J/\psi$& 3872.7& $\phi J/\psi$ & 4116.4&$D^{0}D^{*\pm}_s$ &
3976.7\\
$D^\pm D^{*\mp}$ &3879.5 &$D^{*+}_sD^{*-}_s$& 4223.8& $K^{*\pm}J/\psi$ &
3988.6\\
$\omega J/\psi$& 3879.6 & & &  $K^{*0}J/\psi$ & 3993.0\\
$D^{*0}\bar D^{*0}$ & 4013.6 & & &$D^{*0} D_s^{*\pm}$ & 4118.8   
\end{tabular}
\end{ruledtabular}
\end{table}

\begin{table}
  \caption{Thresholds for open bottom decays.}
  \label{tab:bthr}
\begin{ruledtabular}
\begin{tabular}{cccccc}
Channel& Threshold (MeV)&Channel& Threshold (MeV)&Channel& Threshold
(MeV)\\ 
\hline
$B\bar B$ & 10558& $B_s^+B_s^-$ &10739 & $B B_s$ &10649\\
$B\bar B^*$ & 10604 &$B_s^\pm B_s^{*\mp}$ & 10786 &$B^* B_s$ & 10695\\
$B^*\bar B^*$ & 10650 &$B_s^{*+} B_s^{*-}$ & 10833 &$B^* B_s^*$& 10742  
\end{tabular}
\end{ruledtabular}
\end{table}

The diquark-antidiquark model of heavy tetraquarks predicts
\cite{mppr,bmppr} the existence of a flavour 
$SU(3)$ nonet of states with hidden
charm or beauty ($Q=c,b$): four tetraquarks
($[Qq][\bar Q\bar q]$, $q=u,d$) with neither open 
or hidden strangeness, which have
electric charges 0 or $\pm 1$ and isospin 0 or 1; 
four tetraquarks ($[Qs][\bar Q\bar q]$
and  $[Qq][\bar Q\bar s]$, $q=u,d$) with open strangeness ($S=\pm 1$),
which have electric charges 0 or $\pm 1$ and isospin $\frac12$; 
one tetraquark
($[Qs][\bar Q\bar s]$) with hidden strangeness and zero electric
charge. 
Since in our model we neglect the mass difference of $u$ and
$d$ quarks and electromagnetic interactions, corresponding tetraquarks
will be degenerate in mass. A more 
detailed analysis \cite{mppr}
predicts that such mass differences can be of a few MeV so
that the
isospin invariance is broken for the $[Qq][\bar Q\bar q]$ mass
eigenstates and thus in their strong decays.  
The (non)observation of such states will be a crucial test of the
tetraquark model.

The calculated heavy tetraquark masses are presented in
Tables~\ref{tab:cmass} and \ref{tab:bmass}. The corresponding open
charm  and bottom thresholds are given in
Tables~\ref{tab:cthr} and \ref{tab:bthr}. We find that all $S$-wave
tetraquarks with hidden bottom lie considerably below
open bottom thresholds and thus they should be narrow states which can
be observed experimentally. This prediction significantly differs from the
molecular picture \cite{molec} where bound $B-\bar B^*$ states are
expected to lie very close (only few MeV below) to the corresponding
thresholds. 

The situation  in the hidden charm sector is considerably more
complicated, since most of the tetraquark states are predicted to lie
either above or only slightly below corresponding open charm
thresholds. This difference is the consequence of the fact that the
charm quark mass is substantially smaller than the bottom quark
mass. As a result the binding energies in the
charm sector are
significantly smaller than those in the bottom sector.

\begin{table}
  \caption{Comparison of theoretical predictions for  the masses of
   charm diquark-antidiquark states $cq\bar c\bar q$ (in MeV) and
   possible experimental candidates.}
  \label{tab:cemass}
\begin{ruledtabular}
\begin{tabular}{cccccc}
State&
\multicolumn{3}{l}{\underline{\hspace{3cm}Theory\hspace{3cm}}}
\hspace{-1.5cm}& 
\multicolumn{2}{l}{\underline{\hspace{0.3cm}Experiment
    \cite{exp,belle,babar,d0,cdf,belley,bellex,bellez,babary}
    \hspace{0.3cm}}} 
\hspace{-1.5cm}  \\
$J^{PC}$ &this work & \cite{mppr}&\cite{mppr1} ($cs\bar c\bar s$) 
&state& mass\\
\hline
$1S$\\
$0^{++}$ & 3812 & 3723& & &\\
$1^{++}$ & 3871& 3872$^\dag$& &$X(3872)$ &$3871.2\pm0.4$ \\
$1^{+-}$ & 3871& 3754&& &\\
$0^{++}$& 3852 & 3832&& &\\
$1^{+-}$& 3890 & 3882&& &\\
$2^{++}$& 3968 & 3952&&$Y(3943)$&$3943\pm11\pm13$ \\
$1P$\\
$1^{--}$&4244 & &$4330\pm70$&$Y(4260)$ &$4259\pm8^{+2}_{-6}$
 \end{tabular}
\end{ruledtabular}
\flushleft{${}^\dag$ input}
\end{table}

In Table~\ref{tab:cemass} we compare our results for the charm
diquark-antidiquark bound states with the predictions of Ref.~\cite{mppr}. 
The differences in some of the
mass values can be attributed to the substantial distinctions in
the used approaches. We describe the diquarks as  quark-quark bound
systems dynamically and  calculate their masses, while in
Ref.~\cite{mppr}  they
are treated only phenomenologically. Then we consider the
diquark-antidiquark interaction  and the tetraqark as purely the
diquark-antidiquark composite system.  In distinction Maini et al. 
consider a
hyperfine interaction between all quarks  which, e.g., causes the
splitting of $1^{++}$ and $1^{+-}$ states arising from the $SA$
diquark-antidiquark compositions.  
From Table~\ref{tab:cemass}, where we also give possible experimental
candidates for the neutral tetraquarks with hidden charm,  we see that our
calculation supports  the assumption 
of Ref.~\cite{mppr} that $X(3872)$ can be the axial vector
$1^{++}$ tetraquark state composed from scalar and axial vector
diquark and antidiquark in the relative $S$-wave state.  
On the other hand, in our model 
the lightest scalar $0^{++}$ tetraquark is
predicted to be above the open charm threshold $D\bar D$
and thus to be broad, 
while in the model of Ref.~\cite{mppr} it lies a
few MeV below this threshold, and thus is predicted to be narrow. Our
$2^{++}$ tetraquark also lies higher than the one in Ref.~\cite{mppr},
thus making the
interpretation of this state as $Y(3943)$ less probable but still
compatible with experiment. We
also find that $Y(4260)$ cannot be interpreted as $P$-wave $1^{--}$
state of charm-strange diquark-antidiquark, since its mass is found to
be $\sim 200$ MeV heavier (see Table~\ref{tab:cmass}). A more 
natural
tetraquark interpretation could be the $P$-wave $([cq]_{S=0}[\bar
c\bar q]_{S=0})_{P-{\rm wave}}$ state which mass is predicted in our
model to be 
close to the mass of  $Y(4260)$ (see Table~\ref{tab:cemass}). 
Then the dominant decay mode of $Y(4260)$ would be in $D\bar D$ pairs.     

In summary, we calculated the masses of heavy tetraquarks with hidden
charm and bottom in the diquark-antidiquark picture. In contrast to
previous phenomenological treatments we used the dynamical approach
based on the relativistic quark model. Both diquark and tetraquark
masses were obtained by numerical solution of the quasipotential
equation with the corresponding relativistic 
potentials. The diquark size was also taken into account with the help of
the diquark-gluon form factor in terms of diquark wave
functions. It is important to emphasize  
that, in our analysis, we did not introduce any free adjustable
parameters but used their fixed values from our previous considerations
of heavy and light meson properties. It was found that the $X(3872)$
can be the neutral charm tetraquark state. If it is really a tetraquark,
one more neutral and two charged tetraquark states must exist with close
masses. The ground states of bottom tetraquarks are predicted to have
masses below the 
open bottom threshold and thus should be narrow. The (non)observation of
these states will be an important test of the tetraquark model.

The authors are grateful to V. A. Matveev,  M. M\"uller-Preussker and
V. Savrin for support and useful discussions.  Two of us
(R.N.F. and V.O.G.)  were supported in part by the {\it Deutsche
Forschungsgemeinschaft} under contract Eb 139/2-3 and by the {\it Russian
Foundation for Basic Research} under Grant No.05-02-16243.

\end{document}